\documentclass[twocolumn]{aastex62}
\usepackage{apjfonts} 
\usepackage{amsmath} 
\usepackage{lipsum}

\newcommand{\teff}{$T_{\rm eff}$}

\begin{document}
\title{Measuring the D/H Ratios of Exoplanets and Brown Dwarfs}

\author{Caroline V. Morley}
    \affiliation{Department of Astronomy, University of Texas at Austin, Austin, TX, USA}
\author{Andrew J. Skemer}
    \affiliation{Department of Astronomy \& Astrophysics, University of California Santa Cruz, Santa Cruz, CA, USA}
\author{Brittany E. Miles}
    \affiliation{Department of Astronomy \& Astrophysics, University of California Santa Cruz, Santa Cruz, CA, USA}

\author{Michael R. Line}
    \affiliation{Arizona State University, Tempe, AZ, USA}
\author{Eric D. Lopez}
    \affiliation{NASA Goddard Space Flight Center, Greenbelt, MD, USA}

\author{Matteo Brogi}
    \affil{Department of Physics, University of Warwick, Coventry CV4 7AL, UK}
    \affil{INAF - Osservatorio Astrofisico di Torino, Via Osservatorio 20, 10025, Pino    Torinese, Italy}
    \affil{Centre for Exoplanets and Habitability, University of Warwick, Gibbet Hill Road, Coventry CV4 7AL, UK}

\author{Richard S. Freedman}
    \affiliation{SETI Institute, Mountain View, CA, USA}
    \affiliation{NASA Ames Research Center, Mountain View, CA, USA}

\author{Mark S. Marley}
    \affiliation{NASA Ames Research Center, Mountain View, CA, USA}

\shorttitle{Measuring D/H of Exoplanets}
\shortauthors{Morley et al.}


\begin{abstract}

The relative abundance of deuterium and hydrogen is a potent tracer of planet formation and evolution. Jupiter and Saturn have protosolar atmospheric D/H ratios, a relic of substantial gas accretion from the nebula, while the atmospheres of Neptune and Uranus are enhanced in D by accretion of ices into their envelopes. For terrestrial planets, D/H ratios are used to determine the mechanisms of volatile delivery and subsequent atmosphere loss over the lifetime of the planet. Planets and brown dwarfs more massive than $\sim$13 M$_J$ quickly fuse their initial D reservoir. Here, we simulate spectra for giant exoplanets and brown dwarfs (2 M$_{\rm Neptune}$ to $\sim$10 M$_{\rm Jupiter}$) from \teff=200--1800 K including both CH$_3$D and HDO to determine the observability of these dominant deuterium isotopologues in mid-infrared thermal emission spectra. Colder objects have stronger molecular features in their spectra, due to the temperature-dependence of molecular cross sections. CH$_3$D is easier to observe than HDO at all temperatures considered, due to the strength of its absorption bands and locations of features at wavelengths with few other strong absorption features. We predict that for nearby cool brown dwarfs, the CH$_3$D abundance can be measured at high signal-to-noise with the \emph{James Webb Space Telescope}; for objects from 200--800 K closer than 10 pc, a protosolar D/H ratio would be readily observable in 2.5 hours. Moderately young Jupiter-mass planets (100--300 Myr) and young Neptunes (10 Myr) may be discovered with \emph{JWST} and provide the best targets for detecting deuterium on an exoplanet in the coming decade. Future telescope designs should consider the importance of isotopes for understanding the formation and evolution of planetary atmospheres.

\end{abstract}


\section{Introduction}\label{introduction}

\begin{figure*}
    \centering
    \plotone{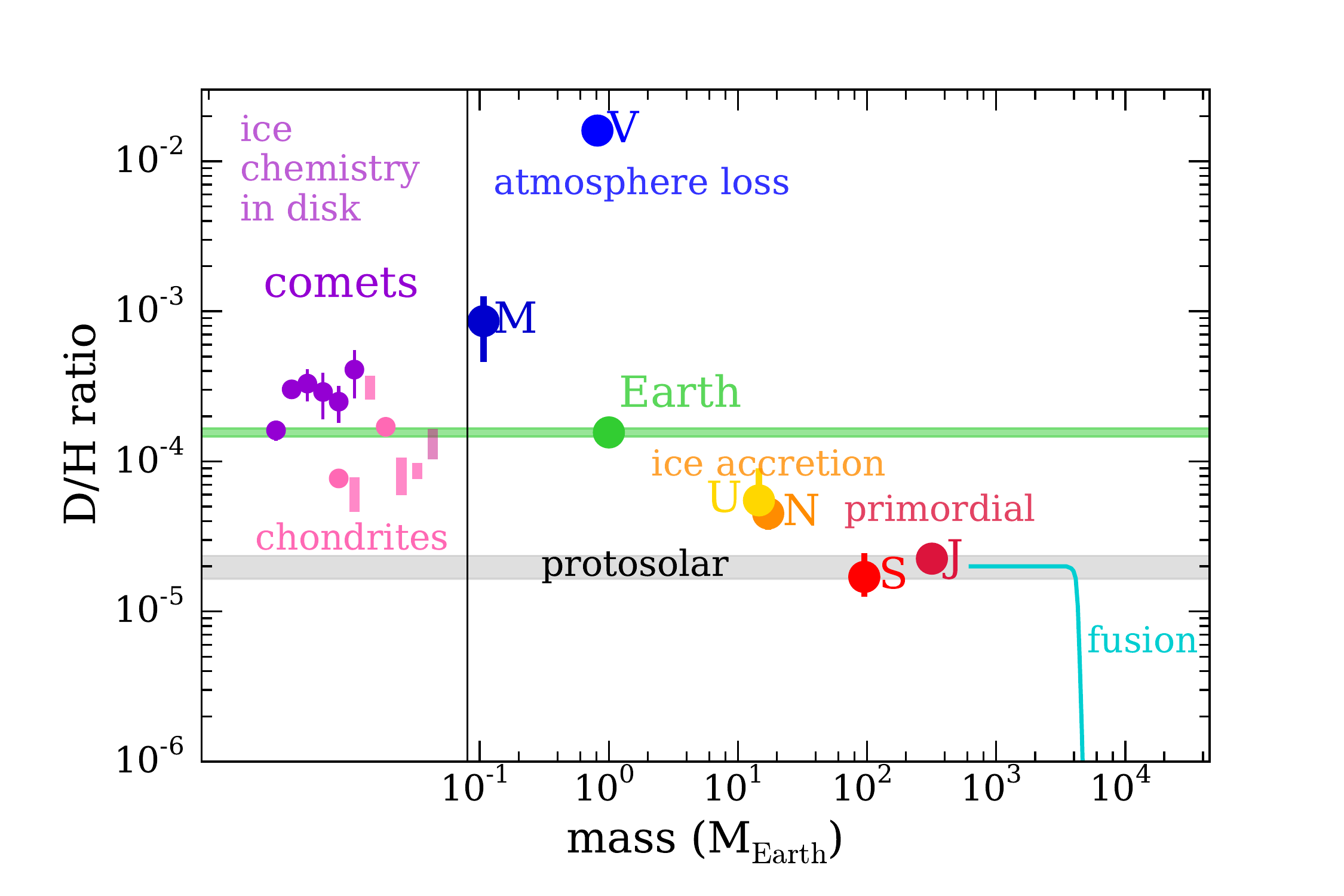}
    \caption{D/H ratio vs. mass of selected solar system objects. From left to right, selected comets, chondrites \& lunar apatite \citep{Hartogh11, Cleeves14}, Mars, Venus, Earth \citep{Drake05}, Uranus, Neptune, Saturn, Jupiter \citep{Hartogh11} and model brown dwarfs \citep{Spiegel11} are shown. The protosolar and Earth D/H ratio are shown as shaded bars. Points with error bars show single measurements; shaded bars show ranges over multiple measurements. }
    \label{dhratios}
\end{figure*}

With current instruments, it is possible to detect molecules and elements in exoplanet and brown dwarf atmospheres \citep[e.g.,][]{Kirkpatrick05, Charbonneau02, Kreidberg14b}. The atmospheric enrichment and ratios of elemental abundances inform us about how an object formed: the atmospheres of brown dwarfs that collapse directly out of a molecular cloud are predicted to form with the same elemental abundances as more massive stars forming in that cloud, while planets that form in a disk via core accretion and gas/planetesimal accretion will have abundance patterns that reflect the material available at their locations in the disk \citep{Oberg11}. For example, planets that form beyond the water ice line will accrete both gas and water-rich ices into their atmospheres; depending on the ratio of gas to ice accreted, the object would have a super- or sub-solar metallicity and C/O ratio \citep{Espinoza17}. 

In studies of the solar system, a complementary and critical tool for tracing planet formation and evolution is the relative abundances of different isotopes of the same element. One important element is deuterium: the deuterium to hydrogen (D/H) ratio traces a host of different physical processes in an atmosphere, including accretion of solids and gas, atmospheric escape, and deuterium fusion \citep{Owen92, Lecluse96}. In this Letter, in Section 2 we will review these processes to provide motivation for the simulations; in Section 3 we describe the model atmospheres; in Section 4, we examine the possibilities for detection of deuterium in an exoplanet atmosphere; and in Section 5 we discuss the implications of such a detection. 

\section{D/H Ratios in Planets and Brown dwarfs}

The D/H ratio can vary substantially from the galactic average; measurements for selected objects are shown in Figure \ref{dhratios}. The primordial D/H ratio in the universe, set by Big Bang Nucleosynthesis, is (2.8 $\pm0.2)\times10^{-5}$ \citep{Pettini08}. The Milky Way's D/H ratio in the gas of the interstellar medium varies along lines of sight by a factor of several, with a mean D/H ratio of (2.0$\pm0.1)\times10^{-5}$, typically measured using absorption line spectroscopy of atomic H and D in the far UV \citep{Prodanovic10}. Ices, both interstellar and within disks, become enhanced in deuterium through several chemical pathways at the cold temperatures (<50 K) present in these environments: gas-phase ion-neutral reactions and grain-surface formation from ionization-generated hydrogen and deuterium atoms from H$_2$ \citep{Cleeves14}. D/H ratios in ices are typically measured at mm wavelengths using sublimating ice (either from comets or around protostars). 

\begin{figure*}
    \centering
    \includegraphics[width=5.5in]{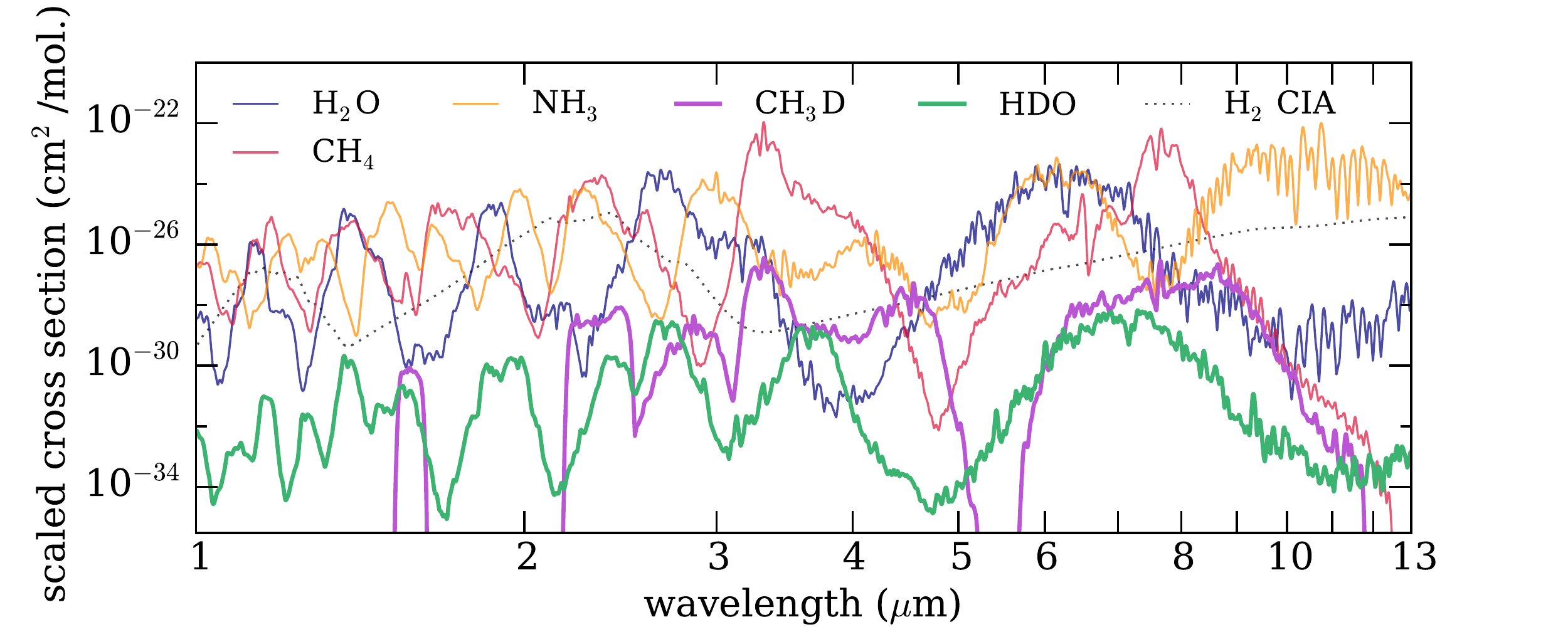}
    \includegraphics[width=5.5in]{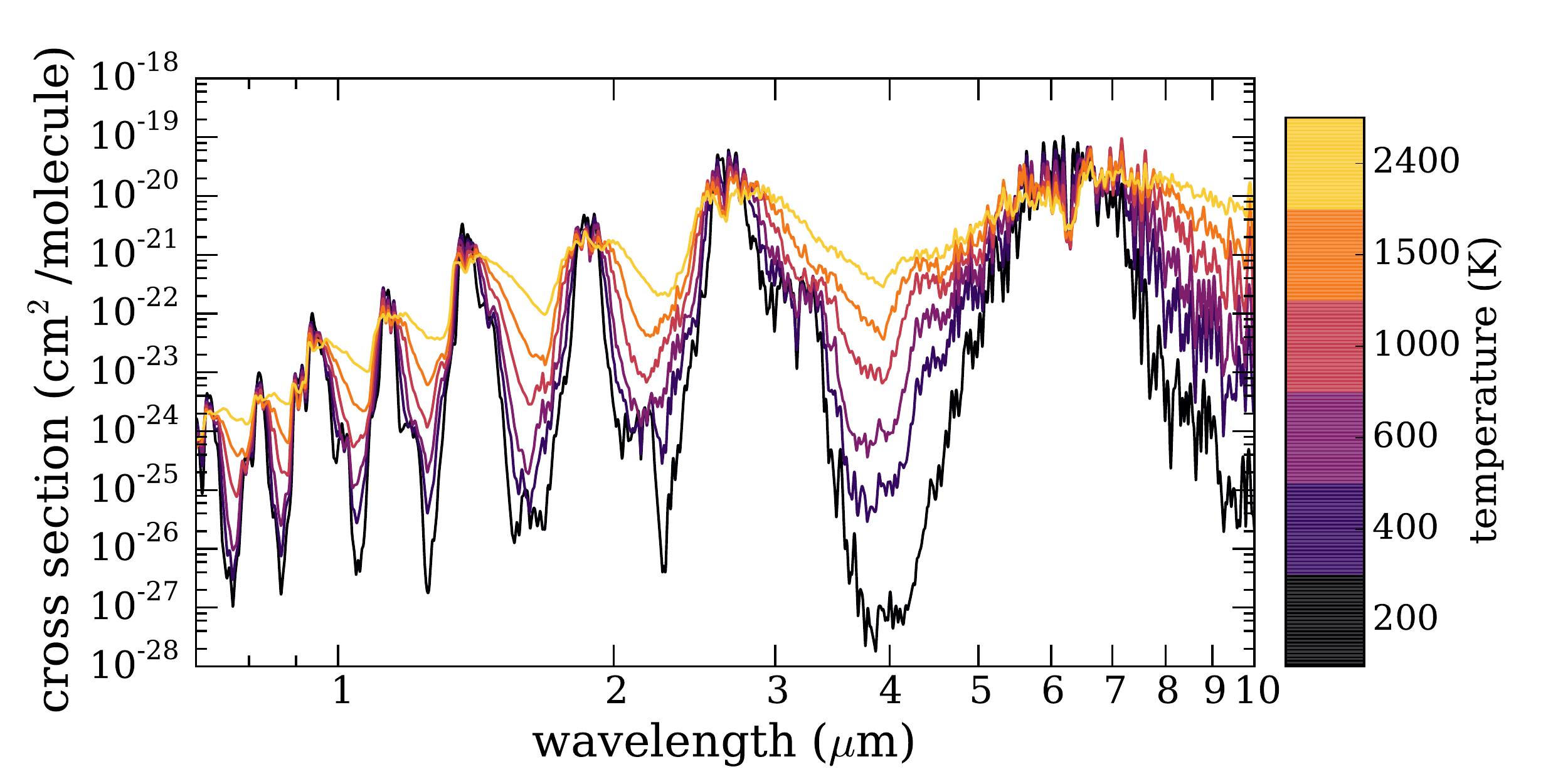}

    \caption{Molecular opacity cross sections for various species. The top panel shows cross sections for  molecules important in cold brown dwarf atmospheres (T=225 K, P=1 bar. Cross sections are scaled by the abundance in chemical equilibrium, assuming D/H= 2$\times10^{-5}$. Bottom panel shows the cross section of H$_2$O at P=1 bar and temperatures from 200 to 2400 K. Colder temperatures lead to larger amplitudes in molecular opacity. }
    \label{opacities}
\end{figure*}

\subsubsection{D/H in Giant Planets Traces Accretion of Solids}

Deuterium was first detected outside of the Earth in Jupiter's atmosphere by \citet{Beer72}. Since then, the deuterium abundances in each of the giant planets' atmospheres have been measured in the near- and mid-infrared using CH$_3$D features and in the visible and far-infrared using HD rotational features \citep{Knacke82, Kunde82,Courtin84, debergh86, debergh90, Feuchtgruber99, Lellouch01, Lellouch10}. Jupiter and Saturn have D/H ratios consistent with the protosolar value, though, intriguingly, different from each other as measured using HD features with Cassini Composite Infrared Spectrometer: Jupiter's D/H is (2.95 $\pm$ 0.55)$\times$10$^{-5}$ and Saturn's is (2.1 $\pm$ 0.13)$\times$10$^{-5}$ \citep{Pierel17}, with Saturn's lower abundance in conflict with predictions from models \citep{Guillot99}.
 Uranus and Neptune's atmospheres are enhanced in deuterium by a factor of $\sim$2.5.

The classical picture is that the giant planets formed by accretion of ices and gases onto a core of $\sim$10--15 M$_{\rm Earth}$ \citep{Stevenson82b}. For Jupiter and Saturn, the relative mass of the core and heavy elements is small compared to the gas accreted, so the D/H ratio is expected to trace the primordial composition of the solar nebula gas. For Uranus and Neptune, more than half of their total masses were accreted as ices; their envelopes are enhanced compared to the solar nebula gas, tracing the relative amount of D-enriched ices that accreted. Assuming that all ices are mainly water, the D/H ratio of the planet is:  

\begin{equation}
    (\rm D/H)_{\rm planet} = (\rm D/H)_{\rm ices} (1-x_{H_2}) + (\rm D/H)_{\rm proto} x_{H_2} 
\end{equation}
where $(\rm D/H)_{\rm planet}$ is the D/H ratio in the planet, $(\rm D/H)_{\rm ices}$ is the D/H ratio in ices, $(\rm D/H)_{\rm proto}$ is the D/H ratio in the protosolar gas, and x$_{H_2}$ is the volume mixing ratio of H$_2$ in the planet \citep[]{Lecluse96}. Much of this ice is incorporated into planetary cores, but models predict that the interior ices exchange deuterium with the hydrogren reservoir \citep{Guillot99}.

\subsubsection{D/H in Terrestrial Planets Traces Both Volatile Accretion and Atmosphere Loss}

Earth, Mars, and Venus have distinct D/H ratios shaped by accretion of volatiles and subsequent atmospheric escape. 

The D/H ratios of comets in the Oort cloud are higher than that of the Earth, suggesting that Earth likely did not accrete the majority of its volatiles from comets \citep{Drake02}, but instead from chondrites. However, \citet{Hartogh11} measured the D/H ratio in a Jupiter-family comet, finding an Earth-like D/H ratio and suggesting that comets and chrondrites may both have played a role in volatile delivery. 

Venus's atmosphere is substantially enhanced in deuterium, by a factor of $\sim$100 above Earth's \citep{Donahue83}. As Venus went through a runaway greenhouse, its oceans evaporated and H$_2$O photodissociated in its upper atmosphere. The relative mass of D relative to H caused H to be more easily lost \citep{Donahue83, ChambHunt}. 

Mars' D/H ratio is similar to Oort cloud comets. Since Mars does not have plate tectonics that recycle the crust with mantle material, this may reflect a late veneer of accretion from comets rather than a primordial reservoir \citep{Drake05}. 

\subsubsection{D/H in Brown Dwarfs Traces Deuterium Fusion}

Brown dwarfs form with approximately protosolar abundances. Those more massive than 20 M$_J$ will efficiently fuse almost all deuterium within 20 Myr; objects under 11 M$_J$ will never fuse their deuterium, retaining primordial abundances \citep{Saumon96, Spiegel11}. The mass at which a brown dwarf will have burned 50\% of its deuterium is roughly 13 M$_J$, but depends slightly on initial He fraction, metallicity, and primordial D/H ratio, and is expected to typically range between 12.2 and 13.7 M$_J$ \citep{Spiegel11}. Most D fusion occurs in the first 100--300 Myr.

\section{Methods}

\subsection{Atmosphere Models}

We model the impact of two D-bearing species long-used to detect deuterium in the solar system, deuterated water (HDO) and deuterated methane (CH$_3$D), on the spectra of free-floating planets and brown dwarfs.  We consider objects below the deuterium-burning limit (M<13 M$_J$). We  include cloud-free objects with effective temperatures from 200 to 1800 K and log g=4.0. These \teff/g pairs cover a broad range of self-luminous free-floating planets. A 4.5 M$_J$ planet will have a log g=4.0 and \teff=200 K at $\sim$5 Gyr; a 10 M$_J$ planet will have a log g=4.0 and \teff=1800 K at $\sim$10 Myr. Radii range from 1.1 R$_J$ (200 K) to 1.2 R$_J$ (800 K) to 1.6 R$_J$ (1800 K). Free-floating planets have been discovered throughout this temperature range \citep[e.g.,][]{Faherty16, Cushing11}. The surface gravity has little effect on the strength of the signal for the small range of surface gravities for objects with M<13 M$_{\rm J}$, so we use a single representative surface gravity. 

Briefly, we calculate temperature structures assuming radiative--convective equilibrium as described in \citet{McKay89, Marley96, Marley99, Saumon08, Morley12, Morley14a}. The opacity database for gases is described in \citet{Freedman14}. The abundances of molecular, atomic, and ionic species are calculated using a modified version of the NASA CEA Gibbs minimization code \citep{McBride92}. We include condensation of cloud species to remove materials from the gas phase, including water and HDO at cold temperatures. Descriptions of recent updates to opacities and chemical equilibrium are described in Marley et al. (in prep.). 

\begin{figure*}[t]
    \centering
    \includegraphics[width=4.9in]{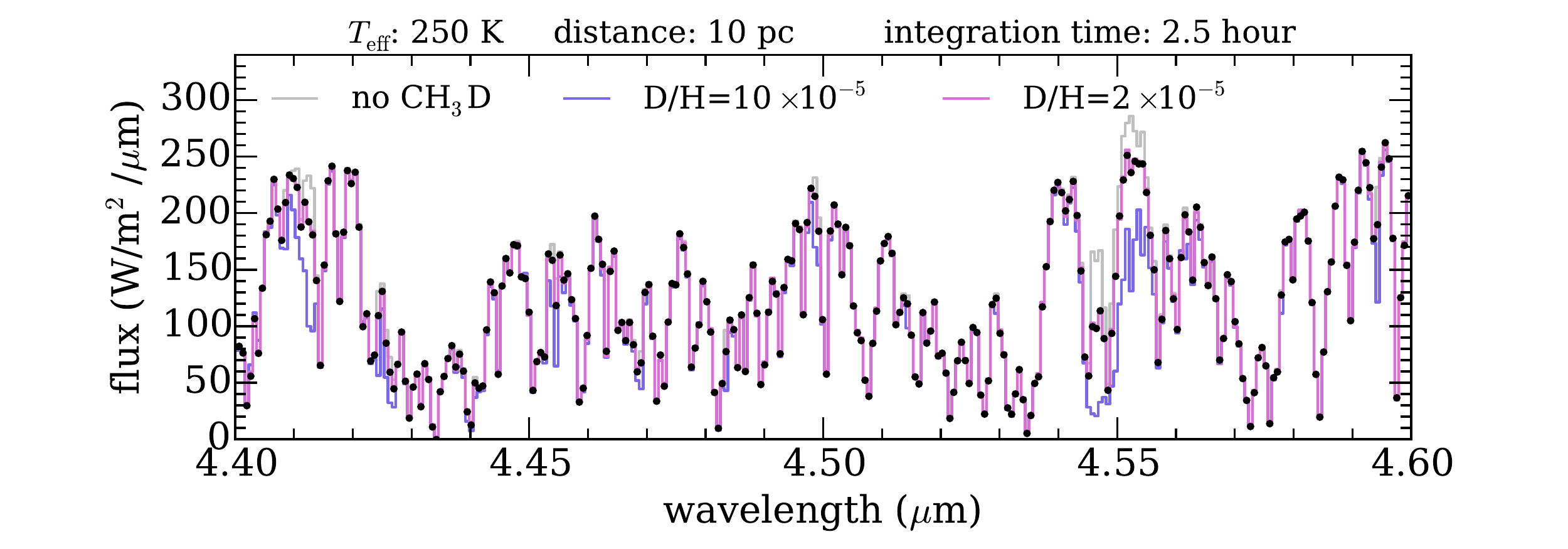}
        \includegraphics[width=4.9in]{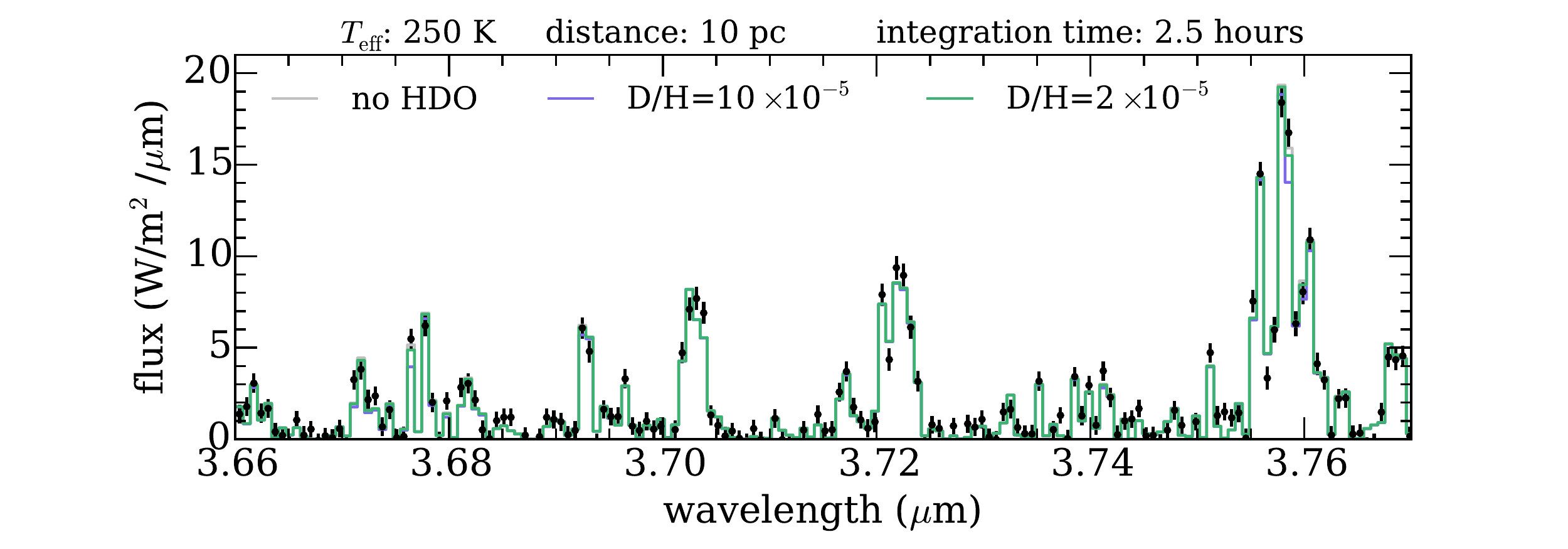}
                \includegraphics[width=4.9in]{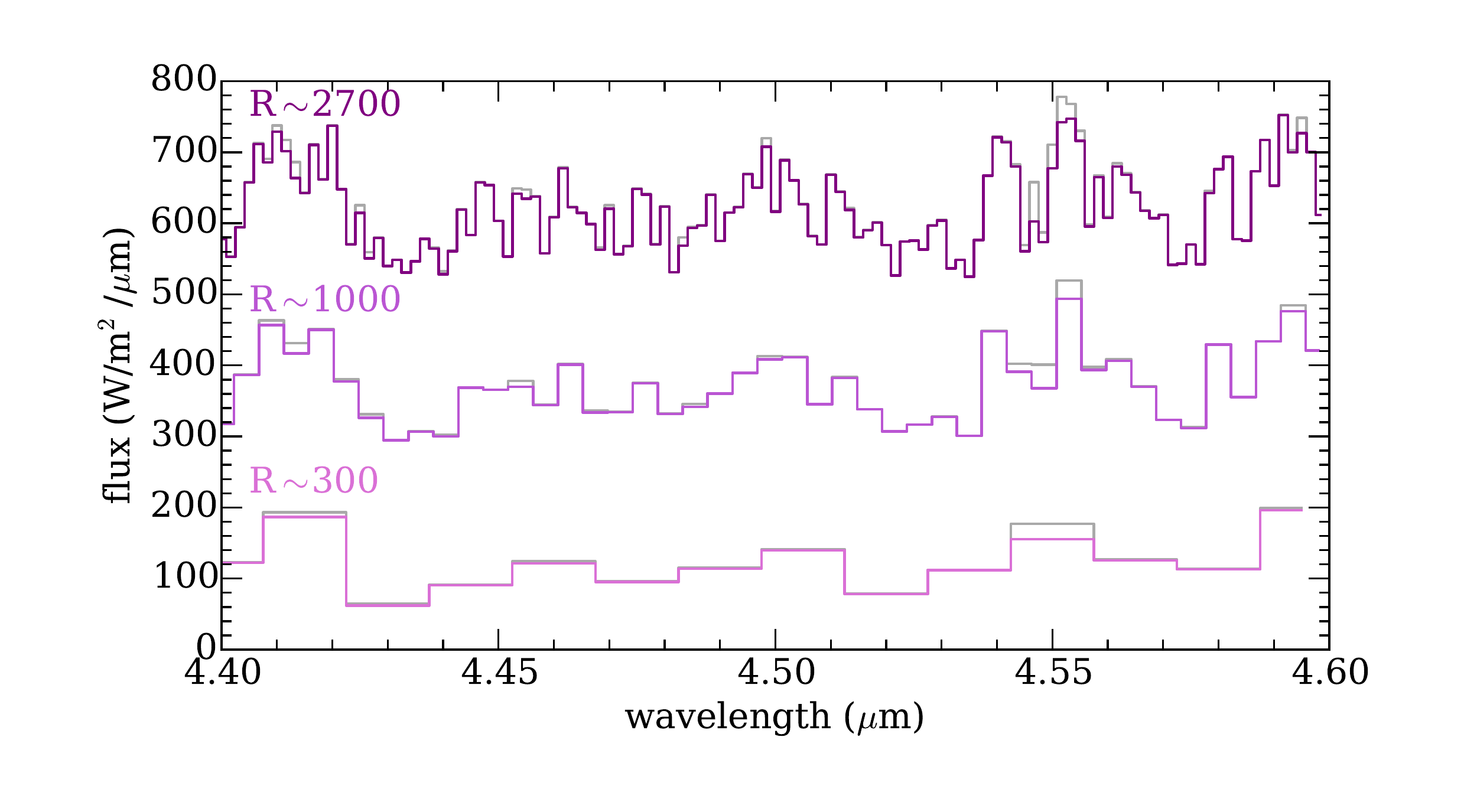}
    \vspace{-0.33in}

    \caption{Model spectra and simulations of JWST data. Top and middle panels show model spectra including CH$_3$D and HDO respectively, for a 250 K object (4--5 M$_{\rm J}$, 2 Gyr). Model spectra include no D, protosolar D/H (2$\times10^{-5}$), and 5$\times$ enhanced D/H (10$\times10^{-5}$). Simulated data for a 2.5 hour observation (assuming a distance of 10 pc) with JWST are shown as black points with error bars. The bottom panel shows how changing spectral resolution changes the CH$_3$D feature for R$\sim$2700, R$\sim$1000, and R$\sim$300. }

    \label{jwst_specs}
\end{figure*}

We calculate moderate-resolution spectra using these pressure--temperature and abundance profiles with the \citet{Morley15} thermal emission code, which uses the open-source radiative transfer code \texttt{disort} \citep{Stamnes88}, which calculates intensities and fluxes in multiple-scattering and emitting layered media using the discrete-ordinate method. 

We include HDO and CH$_3$D opacity when calculating moderate-resolution spectra. Cross sections are calculated using line lists from HITRAN 2012 for CH$_3$D \citep{Rothman13} and ExoMol for HDO \citep{Janca03, Voronin10} and are shown in Figure \ref{opacities}. Our canonical cases assume a protosolar D/H ratio of 2$\times10^{-5}$, typical for formation conditions of stars in the local group; CH$_3$D/CH$_4$ is taken to be 4$\times$D/H in all our simulations and HDO/H$_2$O is 2$\times$D/H, the factors arising from the multiple H atoms in each methane and water molecule.

One of the strongest molecular features of CH$_3$D is at $\sim$4.55\micron. This wavelength region is a ``window'' in atmospheric opacity for cold brown dwarfs, since the important absorbers in the atmosphere---H$_2$O, CH$_4$, and NH$_3$---have low opacity at these wavelengths. For a trace species like CH$_3$D, a smaller amount of the species is necessary for it to be spectroscopically detectable at these window wavelengths because the path length through the atmosphere is longer. In contrast, HDO has features across the near- and mid-infrared (e.g., 3.7 \micron), but these overlap with absorption features from the more abundant species.

\section{Results}

\subsection{CH$_3$D is More Observable than HDO} \label{ch3dobs}

Examples of our model spectra are shown in Figure \ref{jwst_specs}, for a model with \teff=250 K and surface gravity of 4.0, representing a 4--5 M$_{\rm J}$ object with an age of 2 Gyr. 
Here we show the regions with the strongest CH$_3$D and HDO signals respectively, for the temperature with the largest amplitude HDO feature. We find that CH$_3$D has a substantially stronger impact on the spectrum than HDO; this is true at all temperatures studied here. 

To quantify the observability of CH$_3$D and HDO, we simulate the G395H/F290LP grating/filter combination mode of \emph{JWST}/NIRSpec (2.87-5.14 \micron, R$\sim$2700), using the \emph{JWST} online exposure time calculator tool \citep{Pontoppidan16}. We assume that each object is 10 pc away, to match the distances of known cold brown dwarfs (of the known Y dwarfs, 14 of those with measured distances are within 11 pc and 8 more are within 20 \citep{Leggett17}). We assume total observation times of 2.5 hours, including dither time. Dithering is necessary because above SNR$\sim$300, \emph{JWST} is flat-field limited.\footnote{https://jwst-docs.stsci.edu/jwst-exposure-time-calculator-overview/ jwst-etc-residual-flat-field-errors}

While the 4.55 \micron\ band of CH$_3$D is in one of the brightest wavelength region for a cold brown dwarf, the 3.7 \micron\ HDO band is within an absorption feature, so the brown dwarf is intrinsically fainter at those wavelengths. Spectra of cool brown dwarfs can therefore be measured at substantially higher SNR at 4.55 \micron\ than 3.7 \micron. Both the relative size of spectral features and the underlying spectrum of the brown dwarf contribute to making CH$_3$D much easier to detect.

\subsection{Minimum Resolving Power \& Wavelength Ranges Needed }

The molecular bands of CH$_3$D are relatively broad in wavelength (see Figure \ref{opacities}); however, to identify individual features within that broad band, a resolving power of $\sim$1000 is required (see bottom panel of Figure \ref{jwst_specs}). The most useful wavelength range is from 4--5 \micron; CH$_3$D also has a feature at 8.5 \micron, but with higher background and fainter source at longer wavelengths with \emph{JWST}/MIRI, the 4--5 \micron\ region is always favored.

\subsection{Spectrum signal-to-noise for D detection}

We calculate the signal-to-noise of the spectrum necessary to distinguish between a model with and without each deuterated species assuming a protosolar abundance (CH$_3$D/CH$_4$ = 4$\times$D/H = $8\times10^{-5}$; HDO/H$_2$O = 2$\times$D/H = 4$\times10^{-5}$), using a chi-squared rejection test \citep{Gregory}. We note that this is an imperfect method for claiming a `detection', but remains an intuitive way for the reader to picture the relative ease of detection for different simulated spectra. We verify this approach against a full Bayesian retrieval in Section \ref{retrievals}.

The SNR needed generally increases with increasing temperature. HDO always requires a SNR greater than 200 for a 10$\sigma$ detection; CH$_3$D requires a SNR less than 200 for objects with temperatures $\leq$600 K. Water and HDO condensation at \teff<400 K decreases the strength of HDO absorption for the coldest models.

Figure \ref{snr} summarizes our results; in the top two panels, the shaded regions show the SNR achievable in 2.5 hours, and the symbols show the SNR required for a 10-$\sigma$ nominal detection of the deuterated species. We find that HDO is never detectable in our simulated observations in a 2.5 hour integration. CH$_3$D is detectable in objects between 200 and 800 K in <2.5 hrs. The bottom panel shows the fraction of brown dwarfs expected to be less massive than the 13 M$_{\rm J}$ deuterium-burning limit based on a simulated population; cold brown dwarfs are likely to be low-mass and present the best targets for deuterium searches.

\subsection{Temperature Strongly Controls Presence of D-features}

\begin{figure}
    \centering
    \includegraphics[width=3.65in]{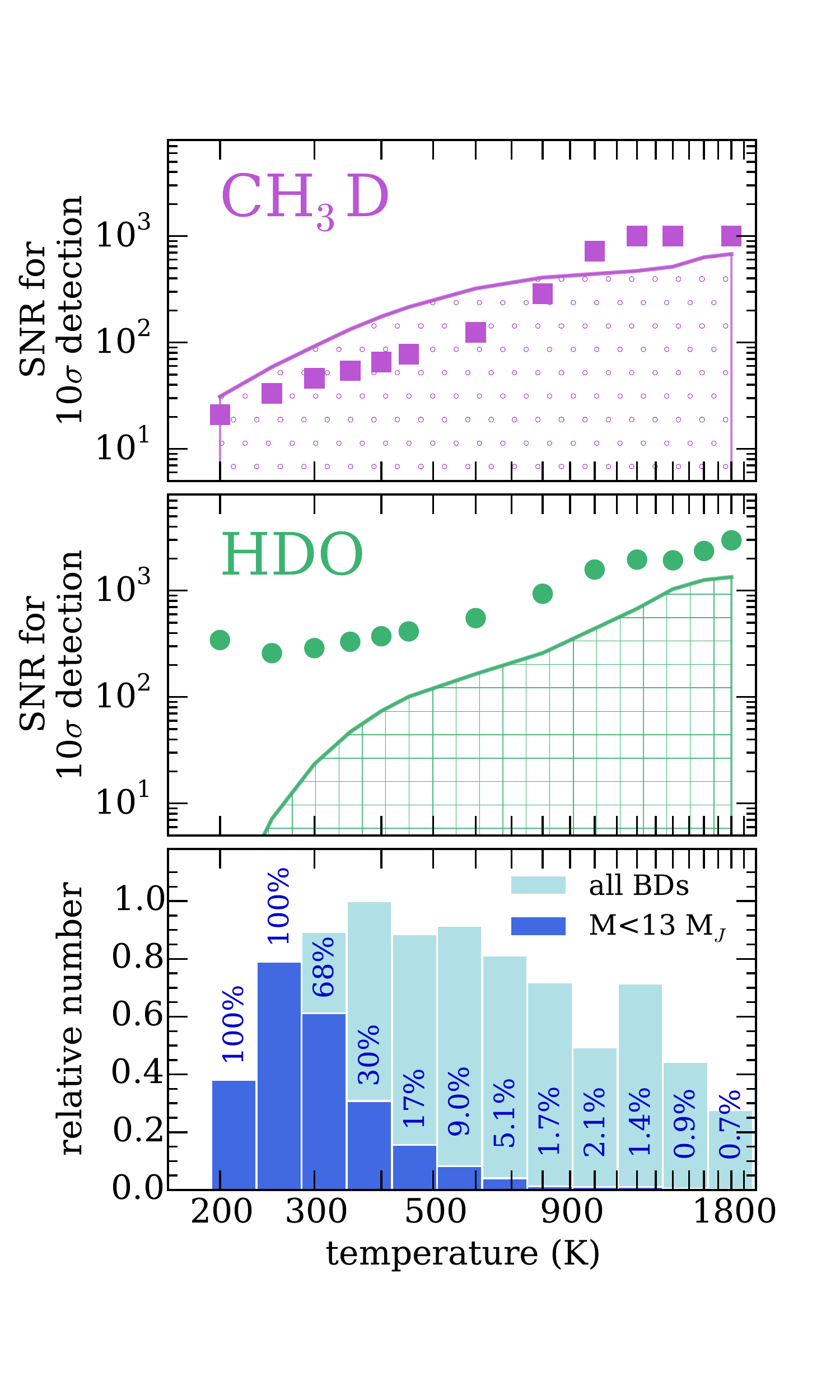}
    \vspace{-15mm}
    \caption{Detectability of deuterated species vs. temperature. The top and middle panels show the signal-to-noise needed for a 10$\sigma$ detection of CH$_3$D and HDO respectively, assuming a spectrum from 3--5\micron\ with R$\sim$2700. The hatched regions show the SNR predicted per spectral element in the region of the spectrum with CH$_3$D or HDO features in a 2.5 hour observation with JWST. CH$_3$D is detectable at high significance for objects cooler than 800 K; HDO is never deteable at high significance in 2.5 hours. Bottom panel shows simulated a brown dwarf population from \citet{Saumon08}; it uses a power law IMF index $\alpha$=1, masses between 0.006 and 0.1 M$_{\rm Sun}$, and uniform age distribution between 0 and 10 Gyr.  }
    \label{snr}
\end{figure}

Figure \ref{snr} clearly shows that \teff\ strongly affects the strength of both CH$_3$D and HDO features; low-temperature objects require lower SNR spectra. This is due to an intrinsic property of molecular opacities: the peak-to-trough amplitude of molecular cross sections in the infrared is strongly correlated with temperature; this is shown for water vapor in Figure \ref{opacities}, bottom panel. At 200 K, the  water cross section between 2.5 and 4.5 \micron\ varies between $\sim10^{-19}$ and $10^{-27}$ cm$^2$/molecule, while at 2000 K it varies between $\sim$ 10$^{-20}$ and $10^{-22}$ cm$^2$/molecule. This striking difference---8 vs. 2 orders of magnitude difference between the absorption band and window---means that the window regions of cold objects probe relatively deeper layers. The larger column of material probed means that a feature from a trace species like CH$_3$D is more prominent in a colder object. This fact has been exploited for decades to detect trace species at $\sim$5 \micron\ in Jupiter, including CH$_3$D, PH$_3$, and GeH$_4$ \citep{Bjoraker86}.

\subsection{Effect of Enhanced Atmospheric Metallicity and D/H Ratio} \label{metallicity}

\begin{figure}
    \centering
    \includegraphics[width=3.5in]{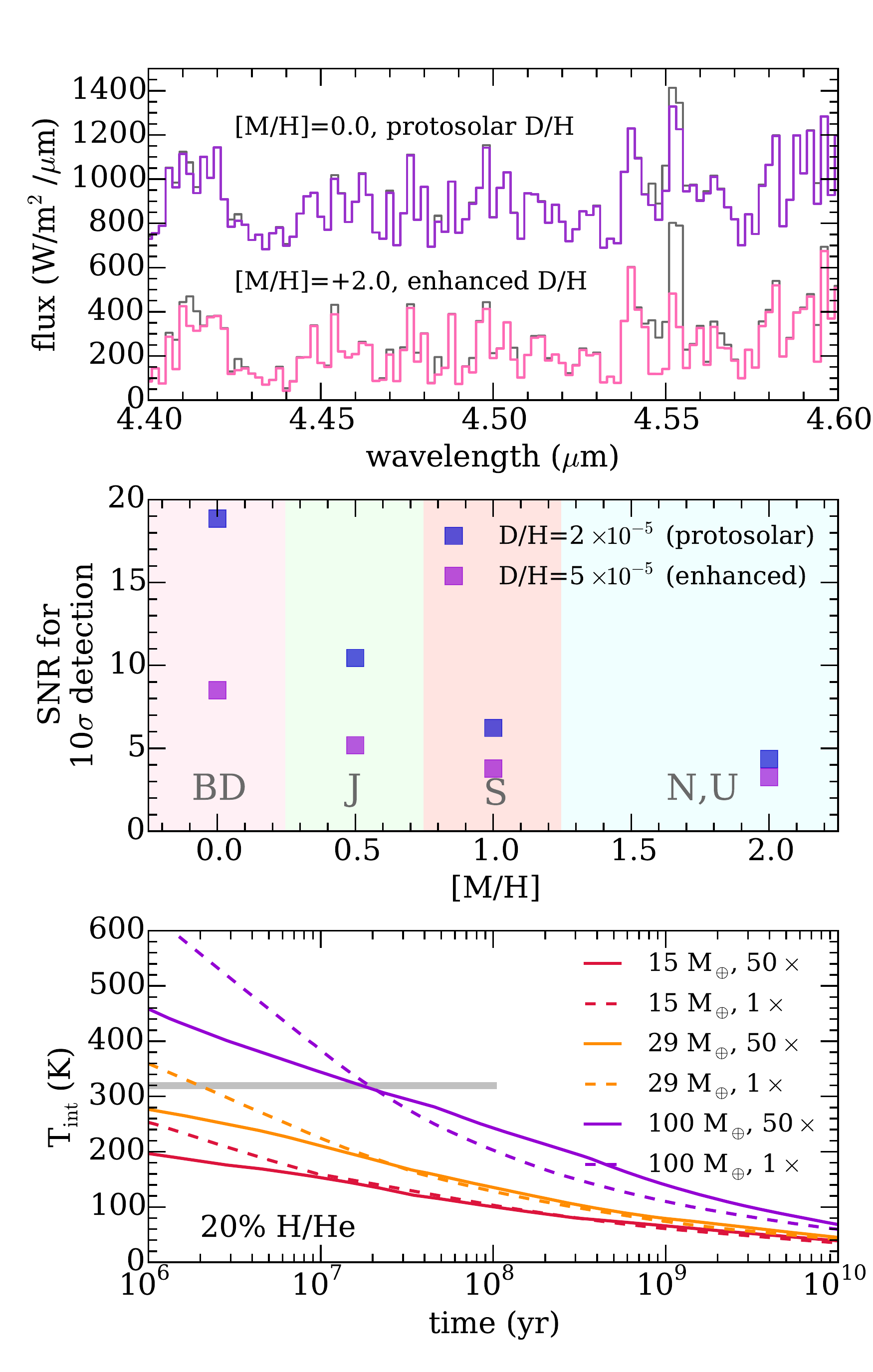}
    \caption{Top panel: model spectra, offset for clarity, of a 320 K, g=15 m/s$^2$, cloud-free object; top model is solar metallicity, with no CH$_3$D (gray line) and a protosolar (2$\times10^{-5}$) D/H ratio (purple line); bottom model is 100$\times$ solar metallicity with no CH$_3$D (gray line) and enhanced (5$\times10^{-5}$) D/H ratio (pink line). Middle panel: SNR per spectral element needed to detect CH$_3$D as a function of metallicity for protosolar and enhanced D/H ratios for a planet with T=320 K, g=15 m/s$^2$. CH$_3$D is easier to detect in high-metallicity objects and for enhanced D/H ratios. Regions are shaded according to the metallicities of local brown dwarfs and solar system giant planets. Bottom panel: intrinsic temperatures of model planets over time for solar and 50$\times$ solar metallicity atmospheric boundary conditions, assuming a 20\% H/He planet by mass. The 320 K temperature modeled above corresponds to a 20 Myr Saturn-mass planet or a 1--2 Myr super-Neptune. }
    \label{fig:neptunes}
\end{figure}

The atmospheres of extrasolar Saturns and Neptunes may be significantly enhanced in metals. We simulate additional spectra with \teff=320 K, chosen to represent a young super-Neptune to Saturn-mass object (30--100 M$_{\rm Earth}$) at 1--20 Myr. We aim to determine how the detectability of CH$_3$D scales with metal-enhancement, using Jupiter-like (3$\times$ solar), Saturn-like (10$\times$ solar) and Neptune-like (100$\times$ solar) metallicities and D/H ratios of 2$\times10^{-5}$ (protosolar) and 5$\times10^{-5}$ (Neptune-like). An example spectrum with high metallicity and enhanced D/H is shown in Figure \ref{fig:neptunes}. 

The strength of the CH$_3$D feature scales with metallicity; increasing the metal-enhancement by a factor of 10 decreases the signal-to-noise needed by a factor of $\sim$3. Similarly, increasing the D/H ratio decreases the SNR needed. These results are summarized in the center panel of Figure \ref{fig:neptunes}. 

The temperature evolution of Neptune and Saturn-mass objects is shown in Figure \ref{fig:neptunes}, calculated using the methodologies described in \citet{Lopez14}, assuming that Neptunes have high initial entropies (`hot start') and cloud-free atmospheres. Young (10 Myr old) Saturn-mass objects are predicted to have temperatures around 400 K, and cool to $\sim$200 K within 100--300 Myr. Neptune-mass planets are colder, with temperatures around 250 K at 10 Myr. Planets in this mass range may be detectable with \emph{JWST} \citep{Beichman10, Schlieder16}.

\subsection{Retrieving D/H Ratios from \emph{JWST} Spectra} \label{retrievals}

Using the atmospheric retrieval tools recently applied to brown dwarfs \citep{Line15, Line17} we quantitatively determine the degree to which we can constrain the CH$_3$D/CH$_4$ ratio with NIRSpec on \emph{JWST} assuming that our models perfectly represent real cold brown dwarfs.  We use the identical forward model parameterization and Bayesian parameter estimation tools to those presented in \citet{Line15, Line17}, but the model is upgraded to include PH$_3$ and CH$_3$D. We tailor the retrieval setup to match the atmospheric structure of a 300 K, logg=5 (cgs), solar metallicity, cloud-free Y dwarf with D/H=2$\times10^{-5}$. We freely fit the abundances of each species individually, pressure--temperature profile, surface gravity, temperature, and radius. 

We find that we can constrain the ratio of CH$_3$D/CH$_4$ to ($8.0\pm0.2)\times10^{-5}$ with a 2.5-hour observation, assuming the same noise models as in Section \ref{ch3dobs}, corresponding to a D/H ratio of (2.0$\pm$0.05)$\times10^{-5}$.  These constraints are remarkable, but in these simulations the forward model is a perfect match to the data; hidden, unforeseen assumptions or systematic errors will inhibit these constraints.

\section{Discussion}

\subsection{Converting CH$_3$D/CH4 to D/H} \label{converting}

Above, we assumed CH$_3$D/CH$_4$ equals 4$\times$D/H (i.e., the molecules are in isotopic equilibrium). However, isotopic exchange depends on temperature. Deep, hotter layers are expected to be in isotopic balance, but colder upper layers probed by these measurements may not be. The actual CH$_3$D/CH$_4$ ratio is determined by a vertical mixing timescale and isotopic exchange timescale. At the top of the atmosphere, the relative abundance $f$ of CH$_3$D/CH$_4$ compared to bulk D/H is 1.25 for Jupiter, 1.38 for Saturn, 1.68 for Uranus, and 1.61 for Neptune \citep{Lecluse96}. The objects considered here are hotter than these planets, and would therefore be closer to isotopic balance ($f$=1.0--1.25). Our approximation is conservative; if $f$ is higher, the amount of CH$_3$D is larger and therefore easier to detect. 

\subsection{Clouds}
Cloud opacity is not included in these calculations. Clouds typically mute features in thermal emission spectra, so if objects of interest are extensively cloudy, these species would be harder to observe. Brown dwarfs with \teff>1200 K have extensive refractory clouds (silicates/iron), while colder objects are relatively well matched by cloud-free models \citep[e.g.,][]{Cushing08} until water clouds form for objects less than 375 K \citep{Morley14a}. Using cold models with thick water ice clouds, we find that cloud-free simulations underestimate the SNR needed to detect CH$_3$D features by $\sim$40--50\%.

\subsection{Interior Physics, Masses, and Ages of the Coldest Brown Dwarfs} 

For objects with known masses, detecting deuterium could allow us to estimate their ages and test models of deuterium fusion as a function of mass. Our calculations demonstrate that deuterium is most observable in the coldest brown dwarfs, the Y dwarfs. 
The bottom panel of Figure \ref{snr} shows a simulation of the number of brown dwarfs at a given temperature, assuming that the initial mass function is a power law (index $\alpha$=1), and with a uniform age distribution between 0 and 10 Gyr \citep{Saumon08}. A 13 M$_J$ object cools to 300 K in 10 Gyr, so for objects under $\sim$300 K, all simulated objects have masses less than the deuterium-burning limit (M<13M$_J$). For objects between 300 and 400 K, $\sim$30--68\% have M<13M$_J$; for hotter objects, the fraction of brown dwarfs with M<13M$_J$ drops off to $\sim$1--2\% for T$_{\rm eff}$>800 K \citep{Saumon96, Spiegel11}. 

Quantifying the presence of deuterium in a range of Y dwarf atmospheres tests the assumptions made in the simulation and thus the properties of field brown dwarfs. Recent work by \citet{Dupuy17} showed that nearby brown dwarfs with mass measurements have systematically younger ages (a median of 1.3 Gyr and an age interval of 0.4--4.2 Gyr for a sample of 10 systems) than the 0-10 Gyr range simulated here; thus, more of them are likely to be deuterium-rich. Sampling a range of 250--500 K objects would test this prediction and provide an independent way of measuring the age of local substellar objects.

\subsection{Detecting Deuterium with High Dispersion Spectroscopy from Ground-based Telescopes}

\citet{Molliere18} recently showed that deuterium may be detectable for exoplanets from the ground using high-dispersion spectroscopy. They find that CH$_3$D may be detectable at 4.7 \micron\ using current instruments for transiting planets below 700 K. Future ELT-class telescopes will allow detections for a wide range of planets. They also find that HDO is more challenging because methane shields HDO absorption. This technique will be complementary to the moderate-resolution spectroscopy with telescopes like \emph{JWST} considered here. 

\subsection{Planet Formation and Envelope Accretion}

In our own solar system, the two largest giant planets, Jupiter and Saturn, have D/H abundances consistent with the protosolar nebula, while the lower mass Neptune and Uranus are enhanced by a factor of several due to accretion of ices during planet formation. Measuring deuterium enhancement or depletion in exoplanets allows us to test planet formation mechanisms in other systems. Multi-planet systems will be particularly valuable since the primordial D/H ratio varies within the galaxy. 

\emph{JWST} will be capable of detecting cool Jupiters and Neptunes around nearby M dwarfs \citep{Beichman10, Schlieder16}. Jupiter-mass exoplanets with ages of 100--300 Myr will have temperatures of 250--300 K \citep{Fortney08b} and radii of 1.1--1.15 R$_J$, so if any such nearby planets are discovered with \emph{JWST}, they will require comparable SNR spectra to the free-floating objects considered here to detect CH$_3$D. 

Young objects (10 Myr) with masses twice that of Neptune are predicted to have temperatures around 200--225 K and radii around 6--9 R$_{\rm Earth}$. If their atmospheres are enhanced in metals and in deuterium, CH$_3$D may be detectable in their spectra (see Section \ref{metallicity}).

\section{Conclusions}

In this Letter, we have presented spectra of model brown dwarfs and free-floating planets including both HDO and CH$_3$D for objects from 200--1800 K. CH$_3$D requires a lower SNR spectrum to detect than HDO at all temperatures. Colder objects have stronger spectral features due to the inherent properties of the cross sections of molecular species, which have larger differences in absorption between troughs and peaks at colder temperatures. For objects from 200--800 K, a protosolar D/H ratio of 2$\times10^{-5}$, would be detectable in spectra with average SNR per spectral element between 20 and 100, readily achievable for these objects with 2.5 hours of integration time with \emph{JWST}, assuming a typical distance of 10 pc. Colder objects will have stronger lines, but require more time to observe since they are fainter; warmer objects have weaker lines and are not readily observable with \emph{JWST}. 

The D/H ratio has been an important tracer of planet formation, gas accretion, and atmosphere evolution in the solar system since the 1970s when CH$_3$D was first detected in Jupiter. For brown dwarfs and free-floating planets in the near future, similar measurements will allow us to map their masses and ages and test models of their interior physics. For exoplanets from Neptune to Jupiter mass, D/H measurements will allow us to understand envelope accretion for planets outside the solar system. These measurements pave the way for future studies of terrestrial planets, for which D/H measurements trace the accretion of their atmospheres and their evolution over their lifetimes. 

\acknowledgements 
 We thank the reviewer for their thoughtful and helpful comments, which have improved the manuscript. We acknowledge the JWST Help Desk, who provided invaluable assistance while developing simulations of JWST observations. This work was performed in part under contract with the Jet Propulsion Laboratory (JPL) funded by NASA through the Sagan Fellowship Program executed by the NASA Exoplanet Science Institute. This research has benefitted from the Y Dwarf Compendium maintained by Michael Cushing at https://sites.google.com/view/ydwarfcompendium/. This research has made use of NASA's Astrophysics Data System.


\end{document}